\documentclass[letter,
               ]{jacow}
\usepackage{pdfpages,multirow,ragged2e} %
\makeatletter%
	\ifboolexpr{bool{xetex}}
	 {\renewcommand{\Gin@extensions}{.pdf,%
	                    .png,.jpg,.bmp,.pict,.tif,.psd,.mac,.sga,.tga,.gif,%
	                    .eps,.ps,%
	                    }}{}
\makeatother

\ifboolexpr{bool{xetex} or bool{luatex}} 
 {}                                      
 {\usepackage[utf8]{inputenc}}           

\usepackage[USenglish]{babel}
\usepackage{amsmath}
\ifboolexpr{bool{jacowbiblatex}}%
 {%
  \addbibresource{jacow-test.bib}
  \addbibresource{biblatex-examples.bib}
 }{}
\begin{document}

\title{Next Generation LLRF Control and Monitoring System for S-Band Linear Accelerators 
\thanks{This work was supported by the U.S. DOE, Office of Science contract DE-AC02-76SF00515.}}

\author{C. Liu\thanks{chaoliu@slac.stanford.edu}, A. Dhar, E. Snively, M. Othman, R. Herbst, E. A. Nanni \\ SLAC National Accelerator Laboratory, Menlo Park, California, USA \\
		}
	
\maketitle

\begin{abstract}
   The low-level RF (LLRF) systems for S-band linear accelerating structures are typically implemented with heterodyne base architectures. We have developed and characterized the next generation LLRF (NG-LLRF) based on the RF system-on-chip (RFSoC) for C-band accelerating structures, and the platform delivered the pulse-to-pulse fluctuation levels considerably better than the requirement of the targeted applications. The NG-LLRF system uses the direct RF sampling technique of the RFSoC, which significantly simplified the architecture compared to the conventional LLRF.  We have extended the frequency range of the NG-LLRF to S-band and experimented with different RFSoC devices and system designs to meet the more stringent requirements for S-band LLRF applications. In this paper, the characterization results of the platform with different system architectures will be summarized and the high-power test results of the NG-LLRF with the S-band accelerating structure in the Next Linear Collider Test Accelerator (NLCTA) test facility at the SLAC National Accelerator Laboratory will be presented and analyzed. 
\end{abstract}

\section{Introduction}
The low-level RF (LLRF) systems of particle accelerators are typically implemented to stabilize the electromagnetic field in accelerating structures. For the RF accelerating structures, the LLRF systems are conventionally based on heterodyne-based architectures. Due to the limited sampling rate of the data converters, the RF signals are mixed up and down from the baseband with analog RF circuits. There are many RF stations for accelerators, and each of the stations has multiple RF inputs and outputs. Therefore, the scale of analog circuits of the LLRF system increases significantly for a large number of RF channels. To meet the design goals of being compact and affordable, we have designed and implemented the next-generation LLRF (NG-LLRF) system based on RF system-on-chip (RFSoC) technology for linear accelerators (LINACs) \cite{liu2024next}, including future collider concepts such as the Cool Cooper Collider \cite{emilio}, which have some of the strictest requirements for LLRF systems. 

We have applied RFSoC technology to a range of high-energy physics and astrophysics instruments operating with RF frequencies up to 6 GHz and performed RF performance characterizations with different RFSoC devices and RFSoC with different configurations \cite{liu2021characterizing ,henderson2022advanced,liuRA,liuRF,liu2023higher,liuSQ}. The development of NG-LLRF at SLAC was focused on C-band LINACs and the prototype demonstrated phase jitter of around 80 fs with the direct loop-back and around 160 fs at the klystron forward coupler with output power up to 16.45 MW \cite{liu:ipac2024-mocn2,liu2025high,liu2025accel}. At SLAC, we have many S-band RF stations, such as the S-band test stands in test laboratories, the S-band station at NLCTA, and the RF gun and approximately 200 klystrons of LCLS \cite{lcls} and FACET, which have obsolete hardware and software at different levels. The RFSoC Gen 3 devices used for NG-LLRF can direct sample RF signals up to 6 GHz and the operation frequencies of the S-band stations are definitely within the range. Therefore, applying the direct sampling scheme of NG-LLRF to the S-band can provide an upgrade path for the LLRF system of existing facilities and a new LLRF platform for future S-band LINAC developments. In this paper, the performance evaluation results of the NG-LLRF for S-band LINACs are summarized and discussed. The initial high-power test results of the NG-LLRF in the S-band with the test stand at NLCTA will be demonstrated and discussed.      

\section{RF Performance Evaluation}

The RFSoC devices integrate RF data converters with the FPGA and processor, enabling the core part of an LLRF system to be implemented in a single chip. The ADCs integrated in RFSoC devices in Gen 3 family have two levels of maximum sampling rate, 2.5 and 5 Giga-sample-per-second (GSPS). For an LLRF system of a LINAC, pulse-to-pulse stability is a critical performance parameter. We have evaluated the phase and magnitude stability with two different RFSoC devices with integrated ADCs that have maximum sample rates of 2.5 and 5 GSPS respectively. The tests were performed with a loopback test circuit for both cases. One of the integrated DACs was programmed to generate an S-band RF pulse with an RF frequency of 2856 MHz and pulse width of 1 \(\mu\)s. The RF signal was looped back to one of the integrated ADCs and the digitized samples were captured. The captured data was post-processed, and the results will be discussed in this section. In this test, the RF pulse rate of 120 Hz was used, which is aligned with the LCLS repetition rate.

\subsection{Pulse-to-pulse Stability with a ZU49DR Device} 

\begin{figure}[!htb]
   \centering
   \includegraphics*[width=1\columnwidth]{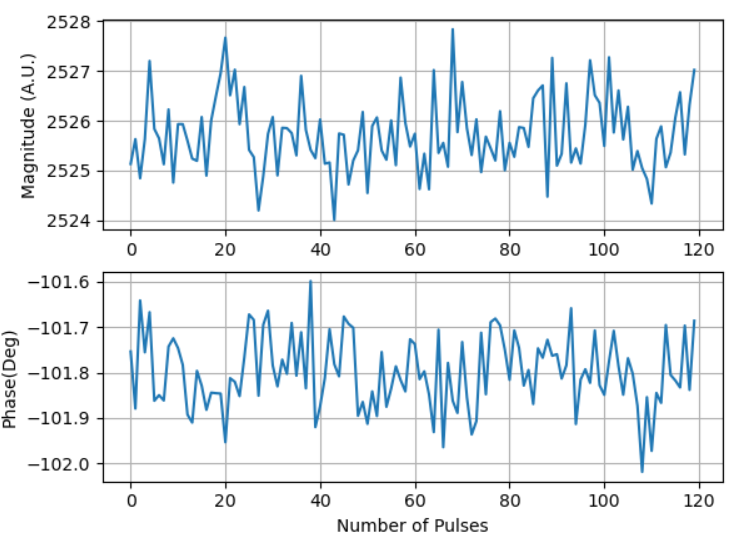}
   \caption{The average magnitude and and phase values of 120 consecutive RF pulses in 1~s with a ZU49DR RFSoC device.}
   \label{fig:zcu216ap}
\end{figure}

The RF performance test was performed with a commercial evaluation board, ZCU216, which carries a ZU49DR RFSoC device. The ZU49DR integrates 16 DACs sampling up to 7 GSPS when the digital up mixer is enabled, and 16 ADCs sampling up to 2.5 GSPS. In this test, a DC pulse was generated in the FPGA firmware at 245.76 MHz data rate and interpolated with a factor of 20 before being up-mixed to 2856 MHz digitally with the integrated numerical controlled oscillator (NCO) in RFSoC. The DAC generates the RF pulse was configured to sample at 4.9152 GSPS. The RF signal looped back to the ADC and was sampled at 2.4576 GHz. The samples were down-mixed at 2856 MHz with the integrated digital mixer and decimated by a factor of 10. Therefore, the bandwidth for this setup is 245.76 MHz, which is significantly higher than the required bandwidth for the LINACs. The downconverted samples were recorded in in-phase (I) and quadrature (Q) format. The data in IQ format was converted to magnitude and phase for further analysis in the software. In this case, the stability of the magnitude and phase in 1~s is analyzed. Within the 1~s time interval, 120 consecutive RF pulses were captured and the average magnitude and phase on the flat top of each pulse were calculated and shown in Figure \ref{fig:zcu216ap}. The fluctuation of magnitude is typically measured by the ratio of the standard deviation of the magnitude values and the average magnitude of all 120 pulses, which is 0.029\% in this case. The fluctuation of the phase measured by the standard deviation of the mean phase values is 0.080\textdegree. 

\subsection{Pulse-to-pulse Stability with a ZU48DR Device}

\begin{figure}[!htb]
   \centering
   \includegraphics*[width=1\columnwidth]{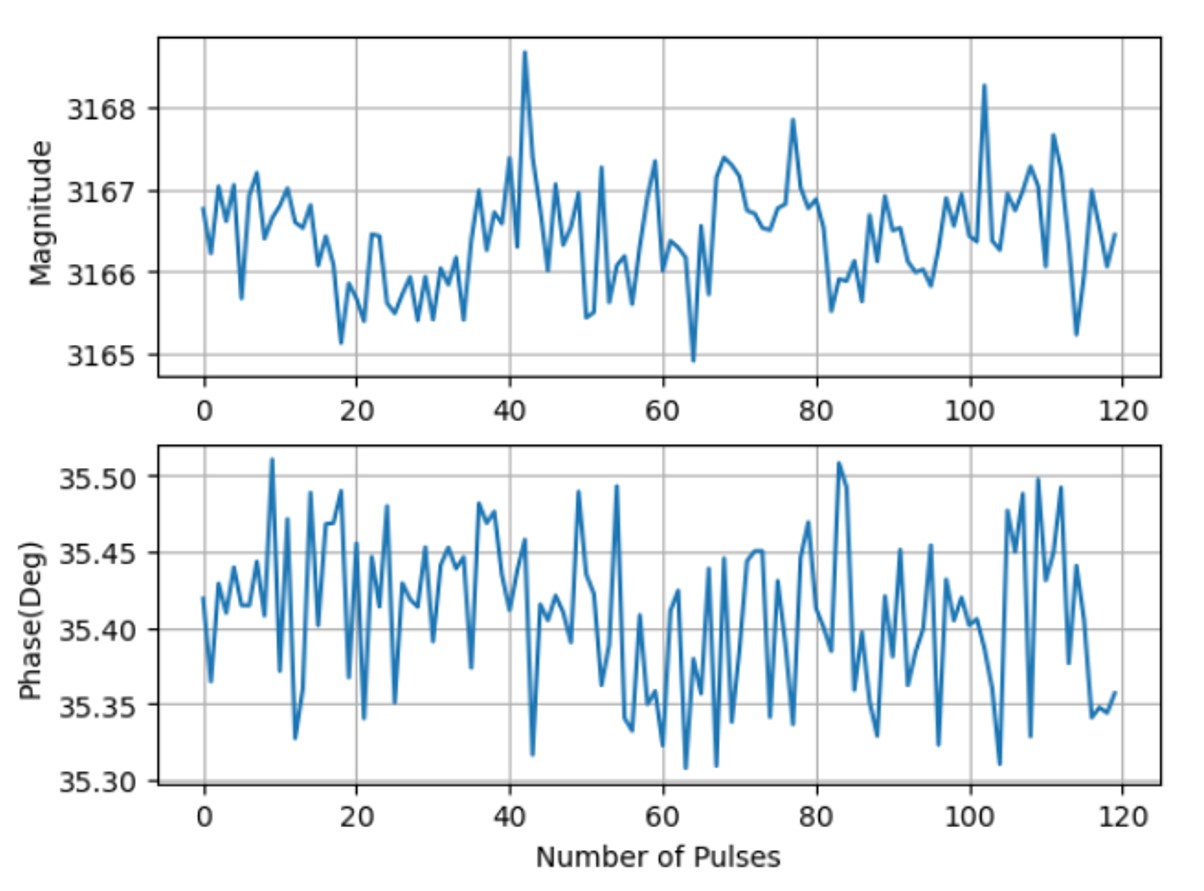}
   \caption{The average magnitude and and phase values of 120 consecutive RF pulses in 1~s with a ZU48DR RFSoC device.}
   \label{fig:zcu208ap}
\end{figure}

\begin{figure}[!htb]
   \centering
   \includegraphics*[width=0.7\columnwidth]{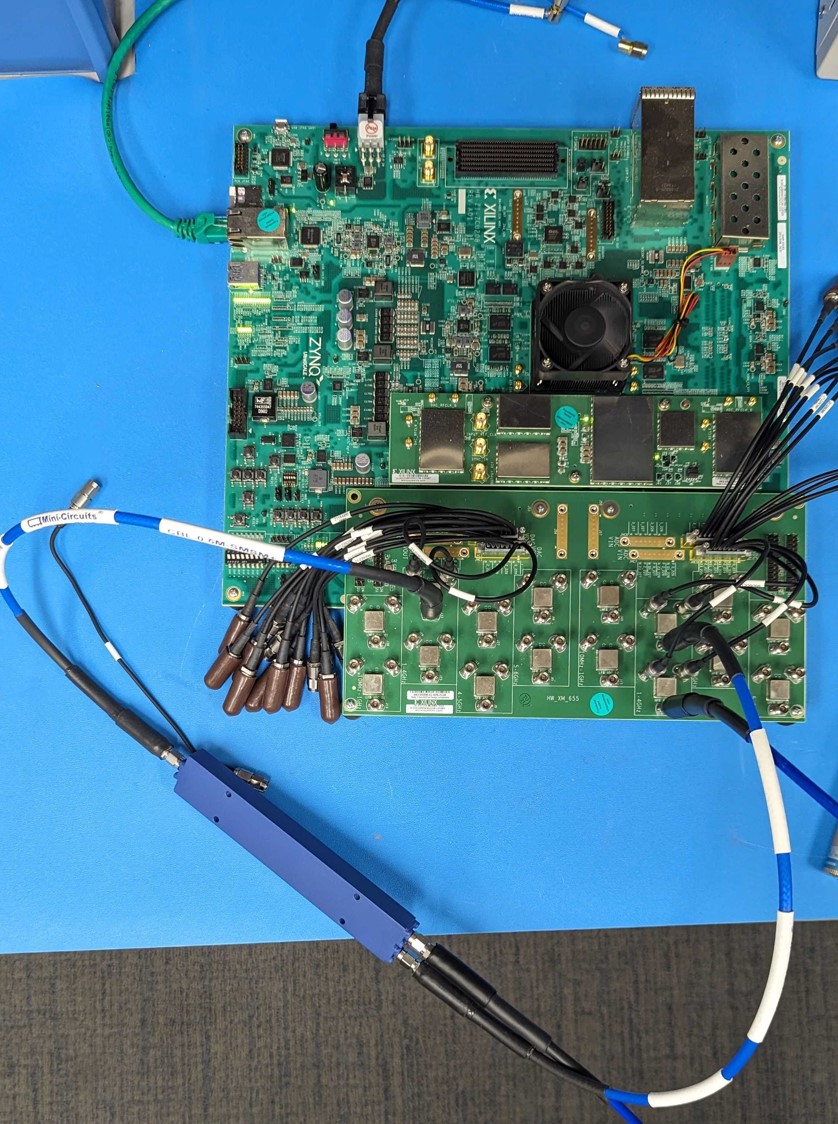}
   \caption{The test setup for the common mode subtraction experiment. }
   \label{fig:zcu208lp}
\end{figure}
The RF performance test was performed with another commercial RFSoC evaluation board, which features a ZU48DR device. Compared with the ZU49DR device, the integrated ADCs in ZU48DR has a maximum sampling rate of 5 GPSP. The test in this case used a similar setup and procedure as the test described in the previous section. However, the ADC was configured to sample at 4.9512 GSPS, which means that the sample rate was doubled compared to the test with the ZU49DR device. Figure \ref{fig:zcu208ap} shows the fluctuation in magnitude and phase in 1s with a ZC48DR device. The magnitude fluctuation is 0.023\% and the phase fluctuation is 0.050\textdegree, which is equivalent to a phase jitter of 48.63 fs. Compared with the fluctuation with the ZU49DR device, the higher sampling rate of the ADC gained approximately 20.7\% and 40.0\% improvements in magnitude and phase fluctuation, respectively. The magnitude and phase fluctuation levels are lower than the lowest short-term stability specifications 0.08\% and 0.07\textdegree of all the LCLS sectors \cite{lcls}.  

\begin{figure}[!htb]
   \centering
   \includegraphics*[width=1\columnwidth]{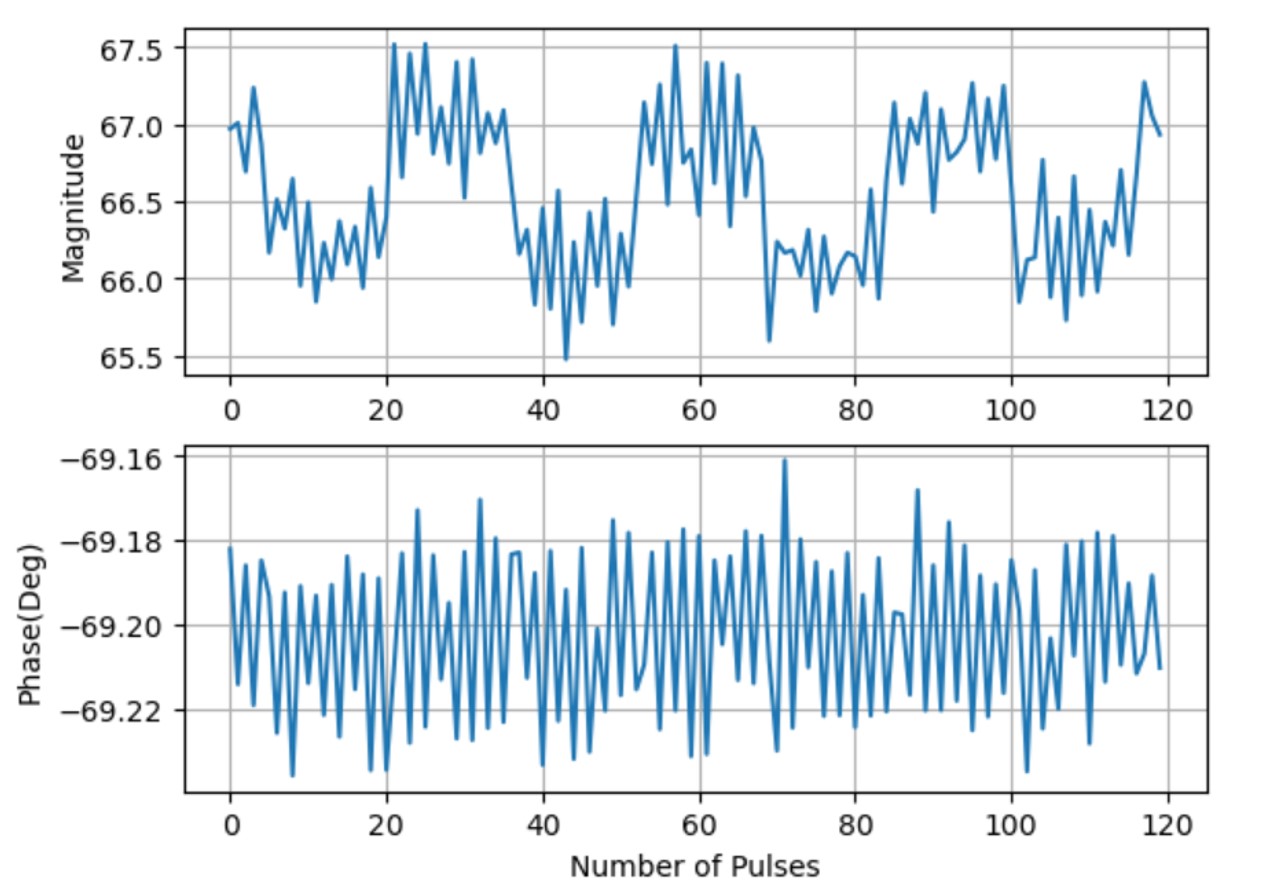}
   \caption{The average magnitude and and phase values after common mode subtraction of 120 consecutive RF pulses in 1~s with a ZU48DR RFSoC device.}
   \label{fig:zcu208sb}
\end{figure}

For the LLRF system implemented with RF mixers, the reference signal is normally down-converted in the same circuits as other RF signals and subtracted from the base-band signals to minimize the local oscillator (LO) noise. For an RFSoC-based LLRF, RF mixing circuit was eliminated and the up and down conversions are fully implemented digitally. However, phase noise may still be introduced by the NCO and the digital mixing process. Therefore, we adapted a similar subtraction technique to reduce noise and the test circuit is shown in Figure \ref{fig:zcu208lp}. The RF signal generated by the integrated DAC in RFSoC is divided into two RF signals, which are looped back to two integrated ADCs in RFSoC via cables with different length to introduce a phase shift. The samples of the two ADCs are subtracted from each other and the fluctuation levels are shown in Figure \ref{fig:zcu208sb}.  The residual magnitude after subtraction is low, but it revealed a low-frequency mismatch and a DC offset between the ADC channels. The phase fluctuation is 0.0196\textdegree, which is equivalent to a phase jitter of 19.06 fs. The pulse-to-pulse phase jitter with the subtraction technique is considerably lower than the 0.07\textdegree \(\) requirement of LCLS and the technique can be used for S-band applications with stringent specifications in phase jitter.

\section{High-power RF Measurement at NLCTA with NG-LLRF}

\begin{figure}[!htb]
   \centering
   \includegraphics*[width=1\columnwidth, angle =270]{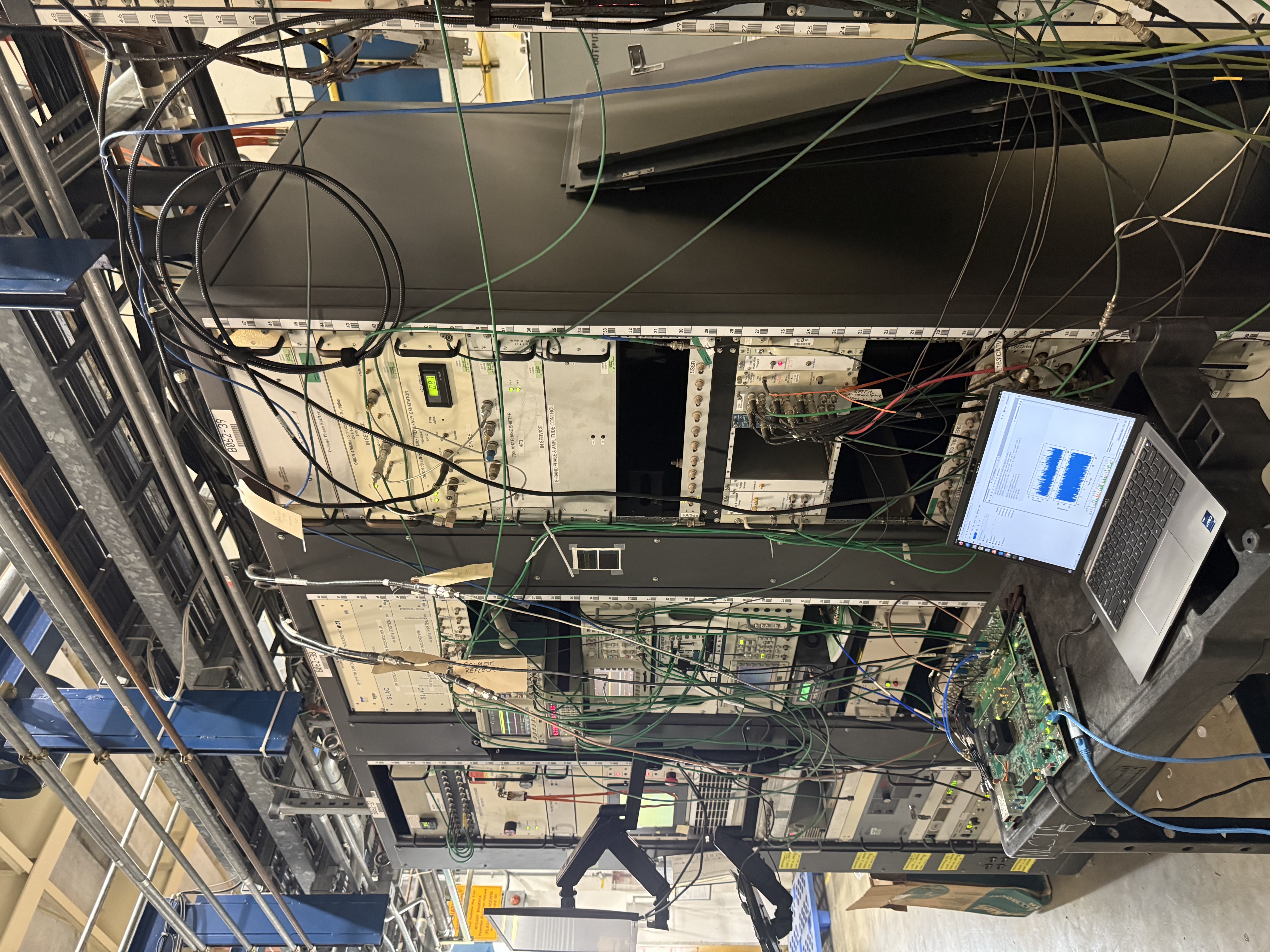}
   \caption{The NG-LLRF interfaced with the S-band test facility at NLCTA.}
   \label{fig:ngllrf}
\end{figure}

\begin{figure}[!htb]
   \centering
   \includegraphics*[width=1\columnwidth]{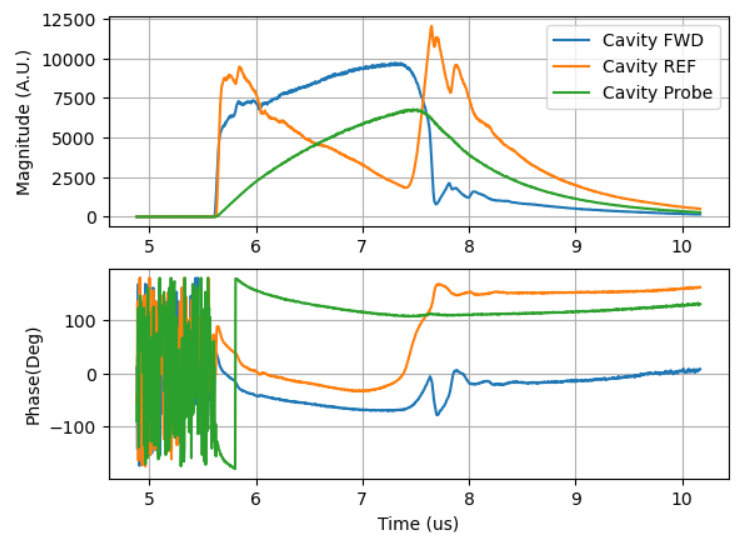}
   \caption{The magnitude and phase of the baseband pulses in the three stages measured by NG-LLRF.}
   \label{fig:ngllrfhp}
\end{figure}

The NG-LLRF system was interfaced with the S-band test stand at NLCTA as shown in Figure \ref{fig:ngllrf} and some preliminary high-power tests were performed. The goal of the test was to verify that the NG-LLRF can measure the RF pulse with reasonable precision. The test stand was driven by the original RF drive system with RF pulses with pulse width around 2 \(\mu\)s at 10 Hz at 2853.8 MHz, which is the cavity resonant frequency in this case. The attenuated RF signals sampled by couplers in three stages, cavity forward (FWD), cavity reflection (REF), and cavity probe, were measured by the RF input channels of NG-LLRF. The RF signals were sampled directly by the integrated in RFSoC. As the down-mixing is fully digital, the conversion frequency can be tuned in software by simply changing the NCO frequency to 2853.8 MHz. The base-band pulses after decimation were then recorded for further analysis. In Figure \ref{fig:ngllrfhp}, the magnitude and phase of the baseband pulses in the three stages are shown. The magnitude of the cavity FWD shown in Figure \ref{fig:ngllrfhp} on top of the pulse increases slowly, which might be the result of the cross coupling between the couplers similar to the one observed in \cite{liu2025high}. The magnitude of the cavity REF decreases linearly as the field fills the cavity. When the RF is switched off, the cavity REF magnitude decays and the phase reverses as the field energy dissipates in the structure. The cavity probe pulse shows the field filling and dissipating directly.       

\section{Summary}

The direct loop-back test of the NG-LLRF prototypes with ZU49DR and ZU48DR demonstrated pulse-to-pulse phase fluctuation of 0.08\textdegree \(\) and 0.05\textdegree, respectively. The ZU48DR-based prototype with the subtraction technique demonstrate 0.0196 \textdegree, which is considerably lower than the LCLS specification. The preliminary results of the high-power test verified that NG-LLRF can measure the RF signals from the accelerating structure with high accuracy. More results will be published with the full S-band NG-LLRF developed. 
%
%
\ifboolexpr{bool{jacowbiblatex}}%
	{\printbibliography}%
	{%
	
	
} 
%
%


\end{document}